\pdfoutput=1
\documentclass[conference]{IEEEtran}
\IEEEoverridecommandlockouts
\usepackage{cite}
\usepackage{amsmath,amssymb,amsfonts}
\usepackage{algorithmic}
\usepackage{graphicx}
\usepackage{textcomp}
\usepackage{algorithm2e}
\usepackage{xcolor}
\def\BibTeX{{\rm B\kern-.05em{\sc i\kern-.025em b}\kern-.08em
    T\kern-.1667em\lower.7ex\hbox{E}\kern-.125emX}}
    
\usepackage[bookmarks=false]{hyperref}
\begin{document}

\title{Slice-Aware Resource Calendaring in Cloud-based Radio Access Networks\\
{\footnotesize \textsuperscript{*}}
}

\author{\IEEEauthorblockN{Zeinab Sasan}
\IEEEauthorblockA{\textit{Computer Engineering Department} \\
\textit{Amirkabir University of Technology}\\
Tehran, Iran \\
z.sasan@aut.ac.ir}
\and
\IEEEauthorblockN{Siavash Khorsandi}
\IEEEauthorblockA{\textit{Computer Engineering Department} \\
\textit{Amirkabir University of Technology}\\
Tehran, Iran \\
khorsandi@aut.ac.ir}
}

\maketitle

\begin{abstract}
Network slicing has been introduced in 5G/6G networks to address the challenge of providing new services with different and sometimes conflicting requirements. With SDN and NFV technologies being used in the design of 5G and 6G wireless network slicing, as well as the centralization of control over these technologies, new services such as resource calendaring can also be used in wireless networks. In bandwidth calendaring, traffic with a low latency sensitivity and a high volume is shifted to later time slots so that applications with a high latency sensitivity can be served instead. 
We discuss how to calendar radio resources in the C-RAN architecture, which also makes use of network slicing. This is referred to as Slice-Aware Radio Resource Calendaring. A model of the problem is developed as an ILP problem and two heuristic algorithms are proposed for solving it due to complexity of optimal solution. Observations have shown that when resources are shared between tenants, the number of accepted requests  increases.

\end{abstract}


\section{Introduction}
One of the major challenges in the next generation of wireless networks is the emergence of new services with different requirements, so that sometimes these requirements are in conflict with each other \cite{vassilaras2017algorithmic}.
The solutions to the this challenge in wireless networks is the introduction of network slicing technology. 

The concept of network slicing enables the creation of multiple logically separate networks on top of a common physical infrastructure platform, allowing a flexible ecosystem to be created to facilitate technical and business innovation, utilizing physical and logical network resources to form a programmable and open multi-tenant network environment \cite{afolabi2018network}.As a result, in architecture based on network slicing, any application can be served according to its requirements. Network slicing enabler technologies include SDN and NFV, which refer to centralized networking capabilities as well as the separation of software from hardware. 

In addition, the Cloud-based RAN (C-RAN) architecture separates the Base Band Units (BBUs) from the Remote Radio Heads (RRHs) in base stations and pools them together in a centralized manner \cite{morais2020sdn}. This Centralization allows mobile operators to effectively allocate resources and reduce operating and capital costs \cite{elias2018radio}.

The move of the wireless networks towards softwarization and centralized management and configuration with SDN and NFV technologies has made it possible to introduce new services in wireless networks. One of these important services is bandwidth calendaring. Bandwidth calendaring refers to shifting some large data transfers with fewer real-time constraints to times when the network is less congested \cite{gkatzikis2016bandwidth}. Applications updates are one example of bandwidth calendaring, such as updates that are not sensitive to latency but generate lots of traffic, and based on calendaring methods, these updates are made during the evening hours. 

Bandwidth calendaring exploits the knowledge or estimation of future arrivals of requests and future demands in an optimal way in the network. Calendaring in wireless networks will optimize the use of resources and improve customer service. The network operator can also maximize revenue and profit by optimizing the use of resources.

In most of the research work done so far, the issue of bandwidth clearance in data center networks has been investigated. In recent years, several research projects have also focused on resource calendaring in wireless networks. But so far, the use of calendaring in next generation wireless networks based on network slicing has not been addressed. To the best of our knowledge, this paper is the ﬁrst to study the Slice-Aware Resource Calendaring.

In this paper, we examine resource calendaring in wireless networks that use the C-RAN architecture and network slicing. In the defined problem, the network operator decides on the acceptance of user requests belonging to tenants and slices, as well as the scheduling and resource allocation to them. There are several tenants with eMBB and eMBBRLLC slices. In the eMBB slice a lot of resources are required, but delays are not a concern. The eMBBRLLC slice requires a lot of resources and is delay sensitive \cite{yazar20206g}. Each of the tenants has also reserved a portion of the operator's resources at the time of contracting.

It is the network operator's intention to serve tenants with different slices so that social welfare is maximized and the maximum number of requests is accepted. In addition, the operator should also try to consider resource constraints, slice and tenant level requirements. We model this scenario as an ILP problem, and due to the complexity of its optimal solution, two heuristic algorithms are proposed. According to numerical results, a heuristic method based on resource sharing results in a higher percentage of requests being accepted. 

This paper is organized as follows. Section II discusses related works. Section III presents the system model. Section IV describes two efficient algorithms to compute sub-optimal yet good solutions for the slice-aware resource calendaring problem. Section V illustrates numerical results and performance evaluation. Finally, Section VI concludes the paper and briefly discusses the future research issues.

\section{Related Works}
During the past, bandwidth calendaring was widely used to move large amounts of data from data centers to WANs, and SDN made it easier. As a result, many of the work on this topic has been carried out using SDN technology in data centers and wide area networks.

The paper \cite{gkatzikis2016bandwidth} describes how a carrier network operator must, at minimum cost, accommodate the demands for predetermined, but time-varying, bandwidth. A few of the demands may be flexible, i.e., can be scheduled during a certain time period. Based on column generation, the authors propose a salable problem decomposition. 

An online version of the bandwidth calendaring is presented in the  paper\cite{dufour2017online}.At the moment of decision making, the amount of resources requested is unknown and information such as offline problems is not fully provided to the operator. Online optimization methods are used to solve this problem.

Calendaring of radio resources in C-RAN is considered in the papers \cite{elias2018radio, morcos2019efficient}. The objective of the network operator is to maximize social welfare and ensure resource requirements and delays per request. It should be noted that in this study, neither network slicing nor tenant and slice concepts were examined and our work is the first to consider calendaring in softwarized wireless networks.

The paper \cite{xiang2021resource} provide an optimization framework that considers several key aspects of the resource allocation problem with cooperating Mobile Edge Computing nodes. Proposed model jointly optimizes (1) the user requests admission decision (2) their scheduling, also called calendaring (3) and routing as well as (4) the decision of which nodes will serve such user requests and (5) the amount of processing and storage capacity reserved on the chosen nodes.

\section{System Model}
In this paper, we consider an operator that manages users requests in various slices of the C-RAN. Each slice and user belongs to a tenant. When concluding a contract with the operator, tenants agreed to the number of resources they reserved. So that the tenant's reserved resources equal $R^k$. Additionally, each tenant has two types of slices of eMBB and eMBBRLLC. First slice requires a lot of resources, but is not sensitive to latency, the second slice requires a lot of resources and is sensitive to latency. In general, eMBB slice requests can be shifted to later time slots. $N$ is equal to the decision interval and depending on the scenario of 5G/6G networks, this timing window can be in the range of milliseconds.

The five fields of a request are as follows:
\begin{itemize}
  \item $t_{0}^{k}$: The arrival time of request $k$.
  \item $r^k$: The number of resources request $k$ requires.
  \item $d^k$: Duration of request $k$.
  \item $u_{n}^{k}$: Request $k$ Utility Function in time slot $n$.
  \item $k_t$: A binary number that indicates that request $k$ belongs to tenant $t$
\end{itemize}

According to the arrival time of each request, a utility function is defined for that request. The function is non-ascending and depends on the type of request. An eMBB request has a utility function that is defined from the time of arrival until the end of its deadline and equals 1. The request must be serviced immediately if it is of the eMBBRLLC type, and the utility function is 1 only when the request arrives. Table \ref{tab1} summarizes the parameters and decision variables.

We use the following sets to model and solve this problem:
\begin{itemize}
  \item $R=\{1,...,R\}$: Set of the resources.
  \item $K=\{1,...,K\}$: Set of the requests.
  \item $N=\{1,...,N\}$: Set of the time slots.
\end{itemize}

\begin{table}[htbp]
\caption{Parameters and Variables Definition}
\begin{center}
\begin{tabular}{|l|cc|}
\hline
\textbf{Parameter} &
  \multicolumn{1}{c|}{\textbf{Category}} &
  \textbf{Definition} \\ \hline
$K$ &
  \multicolumn{1}{c|}{} &
  Total number of requests \\ \cline{1-1} \cline{3-3} 
$R$ &
  \multicolumn{1}{c|}{General} &
  \begin{tabular}[c]{@{}c@{}}Total number of available resource\\  at each time slot\end{tabular} \\ \cline{1-1} \cline{3-3} 
$N$ &
  \multicolumn{1}{c|}{} &
  Total number of available time slots \\ \cline{1-1} \cline{3-3} 
$T$ &
  \multicolumn{1}{c|}{} &
  Total number of Tenants \\ \hline
$U_{t,n}$ &
  \multicolumn{1}{c|}{Tenant} &
  \begin{tabular}[c]{@{}c@{}}Utility function of tenant $t$ in\\  time slot $n$\end{tabular} \\ \cline{1-1} \cline{3-3} 
$R^t$ & 
  \multicolumn{1}{c|}{} &
  \begin{tabular}[c]{@{}c@{}}Number of resource \\  reserved by tenant $t$\end{tabular} \\ \hline
$t_{0}^{k}$ &
  \multicolumn{1}{c|}{} &
  \begin{tabular}[c]{@{}c@{}}Time slot in which the request\\  $k$ arrives\end{tabular} \\ \cline{1-1} \cline{3-3} 
$r^k$ &
  \multicolumn{1}{c|}{} &
  \begin{tabular}[c]{@{}c@{}}Number of resource requested\\  by $k$ \end{tabular} \\ \cline{1-1} \cline{3-3} 
$d^k$ &
  \multicolumn{1}{c|}{Request} &
  Duration of request $k$ \\ \cline{1-1} \cline{3-3} 
$u_{n}^{k}$ &
  \multicolumn{1}{c|}{} &
  \begin{tabular}[c]{@{}c@{}}Utility function of request $k$ starting \\ to be served at time slot $n$\end{tabular} \\ \cline{1-1} \cline{3-3} 
$k_t$ &
  \multicolumn{1}{c|}{} &
  \begin{tabular}[c]{@{}c@{}}Binary parameter that tells if the\\ $k$-th request is belong to tenant $t$\end{tabular} \\ \hline
\multicolumn{1}{|c|}{\textbf{Variable}} &
  \multicolumn{2}{c|}{\textbf{Definition}} \\ \hline
$x_{n}^{k}$ &
  \multicolumn{2}{c|}{\begin{tabular}[c]{@{}c@{}}Binary decision variable that denotes \\ the time slot in which the request $k$ starts.\\  This variable is equal to 1 exactly in one and\\  only one slot, and 0 elsewhere.\end{tabular}} \\ \hline
$r_{n}^{j,k}$ &
  \multicolumn{2}{c|}{\begin{tabular}[c]{@{}c@{}}Binary decision variable that tells if \\ the $j$-th resource is allocated to\\  request $k$ at time slot $n$. This variable\\  is equal to 1 if true, and 0 otherwise.\end{tabular}} \\ \hline
\end{tabular}
\label{tab1}
\end{center}
\end{table} 

This paper is based on the following assumptions:
\begin{itemize}
  \item As soon as a request is initiated, the execution process cannot be interrupted.
  \item Over time, the amount of resources a request requires remains the same.
  \item The decision-making process occurs after all requests have been received and the problem is offline (through forecasting method or previous knowledge).
\end{itemize}

Using Integer Linear Programming (ILP), we formulate the optimal calendaring problem. A C-RAN operator allocates resources with the objective of maximizing social welfare while considering resource constraints, delays, and the resources reserved for each tenant. In the case of the request $k$ being accepted at time slot $n$, the decision variable $x_n^k$ is 1, otherwise 0. The decision variable $r_n^{j,k}$ will be 1 if resource $j$ is assigned to request $k$ at time slot $n$, otherwise it will be 0.

\begin{equation}
Maximize  \sum_{t \in T, k \in K, n \in N: n \geq t_{0}^{k}} k_{t} x_{n}^{k} u_{n}^{k}
\label{eq1}
\end{equation}

\begin{equation}
\sum_{n=t_{0}^{k}}^{N-{d^k}-1} x_{n}^{k} \leq 1 \qquad \forall k \in K\label{eq2}
\end{equation}

\begin{equation}
\sum_{k \in K, n \geq {t_{0}^{k}}} r_{n}^{j,k} \leq 1 \qquad \forall n \in N, j \in R
\label{eq3}
\end{equation}

\begin{equation}
\sum_{ \tau  \in N: \tau  \leq min  \big\{{n, d^{k}}\big\}} x_{n-\tau+1}^{k} \leq \sum_{j \in R} r_{n}^{j,k} \qquad	\forall n \in N, k \in K\label{eq4}
\end{equation}

\begin{equation}
\sum_{j \in R} r_{n}^{j,k} = {r^k}  \big( \sum_{\tau \in N: \tau  \leq min  \big\{{n, d^{k}}\big\}} x_{n-\tau+1}^{k}  \big) \qquad	 \forall k \in K, \forall n \in N\label{eq5}
\end{equation}

\begin{equation}
\sum_{k \in K, j \in R} r_{n}^{j,k}  \leq R  \qquad \forall n \in N\label{eq6}
\end{equation}

\begin{equation}
\sum_{k \in K, j \in R} k_t r_{n}^{j,k}   \geq  R^t  \qquad	 \forall n \in N, \forall t \in T\label{eq7}
\end{equation}

\begin{equation}
x_{n}^{k}, r_{n}^{j,k}  \in  \big\{0, 1\big\} \qquad \forall n \in N, \forall k \in K, \forall j \in R\label{eq8} \end{equation}

All requests must be accepted before the deadline is completed or rejected (\ref{eq2} Admission Control functionality). Each resource in each time slot should be allocated to a maximum of one request (\ref{eq3}). The resources required for a request must be guaranteed when servicing that request (\ref{eq4} and \ref{eq5}). Resources allocated should not exceed operator's available resources (\ref{eq6}). The sum of resources allocated to each tenant's requests must be more or equal the reserved resources (\ref{eq7}). Lastly, Decision variables are binary (\ref{eq8}).

\section{Slice-Aware Calendaring in C-RAN: Heuristics Algorithms}
We propose heuristic solution in this section due to the difficulty of solving the exact ILP optimization problem. The main approach in the proposed solution is to share resources between different tenants. In other words, when a tenant doesn't need reserved resources, the operator can provide unused resources to other tenants. To better compare the proposed solution based on sharing, the method of dedicated resource allocation is also presented. In the following sections, both heuristic methods are described and in the next section, and in the next section, the function of these two methods is compared.

\subsection{Dedicated Resource Allocation (DRA)}\label{AA}
According to this solution, first of all, the reserved resources are assigned to each tenant separately. Reserved resources are then assigned to each of the tenant's requests. First, all requests that are not accepted are sorted by priority. A request with a closer deadline and more requested resources has a higher priority. If the resources required for the request are available during the $d^k$ period after the start of the time slot $n$, this request will be accepted and otherwise rejected. Algorithm 1 shows the algorithm for this solution.

\RestyleAlgo{ruled}
\begin{algorithm}
\caption{Dedicated Resource Allocation (DRA)}
\For{All Tenants $T$}{
Allocate Dedicated Resources to Tenant $t$\;
Find the Requests of Tenant $t$\;
\For{All Time Slots $N$}{
Compute Priority for Not-Accepted Requests of Tenant $t$ in Time Slot $n$\;
Sort Priority-List of Tenant $t$ \;
\For{All Requests in Priority-List}{
\If{Tenant $t$ Reserverd Resources $R^t$ is Insufficient for Acceptance of Request $k$}
{
    Reject Request $k$\;
}
Find Free Resource for Request $k$ in range $n$:$n$+$d^k$\;
\If{ \# of Candidate Resources is Sufficient for Acceptance of Request $k$}
{
    Request $k$ Accepted and $x_{n}^{k}$=1\;
    \For{$r$ in range(0,$r^k$)}{
    \For{$n'$ in  range($n$, $n+d^k$)}{
    $r_{n'}^k=1$\;
    }
    }
}
\If{Request $k$ Accepted}
{
    Tenant $t$ Resource-Usage ++\;
}
}
}
}
\end{algorithm}

\subsection{Sharing-based Resource Allocation (SRA)}
This method does not allocate resources precisely to each tenant. The unaccepted requests in each time slot are sorted by the priority. A request with a closer deadline and more requested resources has a higher priority. A check is performed on each request in this sorted list to determine whether the resources reserved for the tenant can handle the request. The availability of the required resources during the service period of request will be checked if the tenant has enough reserved resources. The request will not be accepted if there are no free resources during the service period.

The tenant's reserved resources may not be sufficient for this request, so other tenants' resources will be considered. In other words, tenants can share their resources in this method. This request can be accepted if the other tenant has sufficient and free resources. The algorithm for this solution is shown in Algorithm 2.

\begin{algorithm}
\caption{Sharing-based Resource Allocation (SRA)}

\For{All Time Slots $N$}
{
Compute Priority for All of Not-Accepted Requests\;
\For{All Requests in Priority List}
  { Find Tenant $t$ of Request $k$\;
   \eIf{$R^t+r^k$ $>$ Tenant $t$ Reserved-Resources}
   {
    Find Tenant $t'$ with Not-Fulled Reserved-Resources\;
    Find Free Resource for Request $k$ in range $n$:$n+d^k$\;
    \If{\# of Candidate Resources is sufficient for Request $k$}
      {Request $k$ Accepted and $x_{n}^{k}$=1\;
        \For{$r$ in range(0,$r^k$)}{
            \For{$n'$ in range($n$,$n+d^k$)}{
              $r_{n'}^k=1$\;   }    } }

    \If{Request $k$ Accepted}
       { Tenant $t$ Resource-Usage ++\; }
   }
   {
       Find free Resource for Request $k$ in Range $n$:$n+d^k$\;
       \If{\# of Candidate Resources is Sufficient for Request $k$}
          { Request $k$ Accepted and $x_{n}^{k}$=1\;
           \For{ $r$ in range(0,$r^k$)}{
             \For{$n'$ in range($n$,$n+d^k$)}{
                 $r_{n'}^k=1$\; } }
           }      
   \If{Request $k$ Accepted}
      { Tenant $t$ Resource-Usage ++\; }
      
    }
}}
\end{algorithm}

\section{Performance Evaluation}

In this section, the proposed solutions are evaluated for their efficiency. As mentioned in the previous section, the main approach in the proposed solution is based on resource sharing. Dedicated Resource Allocation (DRA) is also presented and their performance are compared with each other. On the Google Colab platform, these algorithms have been implemented and evaluated. In this implementation, the requests are distributed among the 3 tenants uniformly, and each tenant's requests are of the eMBB type with 0.5 probability and eMBBRLLC type with 0.5 probability. Table \ref{tab2} presents the other parameters used in the simulation.

\begin{table}[htbp]
\caption{Settings and Parameters}
\begin{center}
\begin{tabular}{|lcc|}
\hline
\multicolumn{1}{|l|}{\textbf{Parameter}}                                  & \multicolumn{1}{c|}{(a)}              & (b)              \\ \hline
\multicolumn{1}{|l|}{Number of time slots}                                & \multicolumn{1}{c|}{$N$ = 10}         & $N$ = 10         \\ \hline
\multicolumn{1}{|l|}{Number of tenants}                                   & \multicolumn{1}{c|}{$T$=3}            & $T$=3            \\ \hline
\multicolumn{1}{|l|}{Number of resources available in a time slot}        & \multicolumn{1}{c|}{$R$ = 20}         & $R$={[}0; 100{]} \\ \hline
\multicolumn{1}{|l|}{Number of requests}                                  & \multicolumn{1}{c|}{$K$={[}0; 100{]}} & $K$ = 50         \\ \hline
\multicolumn{1}{|l|}{Percentage of eMBB requests}                    & \multicolumn{1}{c|}{50\%}             & 50\%             \\ \hline
\multicolumn{3}{|c|}{\textbf{Per Request Parameters}}                                                                     \\ \hline
\multicolumn{1}{|l|}{Arriving time $t_{0}^{k}$ generated from}            & \multicolumn{2}{c|}{$u.d \in  {[}1: 5{]}$}                    \\ \hline
\multicolumn{1}{|l|}{\begin{tabular}[c]{@{}c@{}}Number of resources requested by request $k$ \\generated  from\end{tabular}} & \multicolumn{2}{c|}{$u.d \in {[}1: 5{]}$}                    \\ \hline
\multicolumn{1}{|l|}{Request duration $d^k$ generated from}               & \multicolumn{2}{c|}{$u.d \in {[}1: 5{]}$}                    \\ \hline
\multicolumn{1}{|l|}{First Tenant Share of Resources}                     & \multicolumn{2}{c|}{0.2}                                 \\ \hline
\multicolumn{1}{|l|}{Second Tenant Share of Resources}                    & \multicolumn{2}{c|}{0.2}                                 \\ \hline
\multicolumn{1}{|l|}{Third Tenant Share of Resources}                     & \multicolumn{2}{c|}{0.6}                                 \\ \hline
\end{tabular}
\label{tab2}
\end{center}
\end{table}

\begin{figure}[htbp]
\centerline{\includegraphics[width=1.0\linewidth]{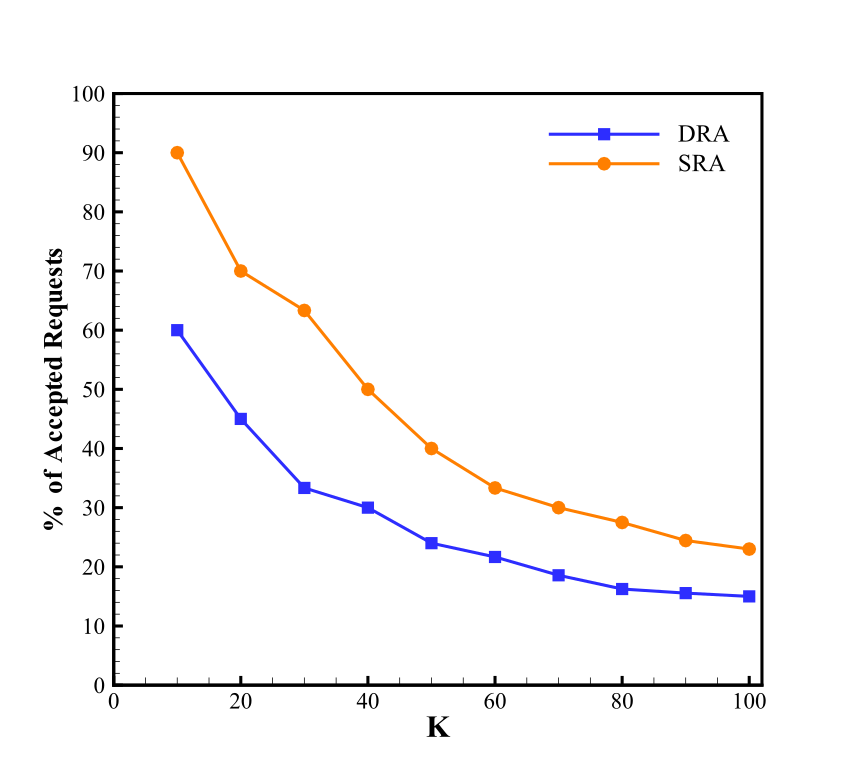}}
\vspace*{-9mm}\hspace*{0mm}\caption{Percentage of Accepted Requests for Constant Resources (a)}
\label{fig1}
\end{figure}

The number of requests was changed first. As shown in Figure \ref{fig1}, The number of requests accepted by the SRA solution is greater than the number of requests accepted by the DRA solution since the sharing method utilizes the resources available to other tenants. 

It is important to note in this section that in general, the sharing method is more efficient than the deductible approach. Nevertheless, as the number of requests increases, the sharing method becomes less efficient and the results of the two algorithms become more similar.

\begin{figure}[htbp]
\centerline{\includegraphics[width=1.0\linewidth]{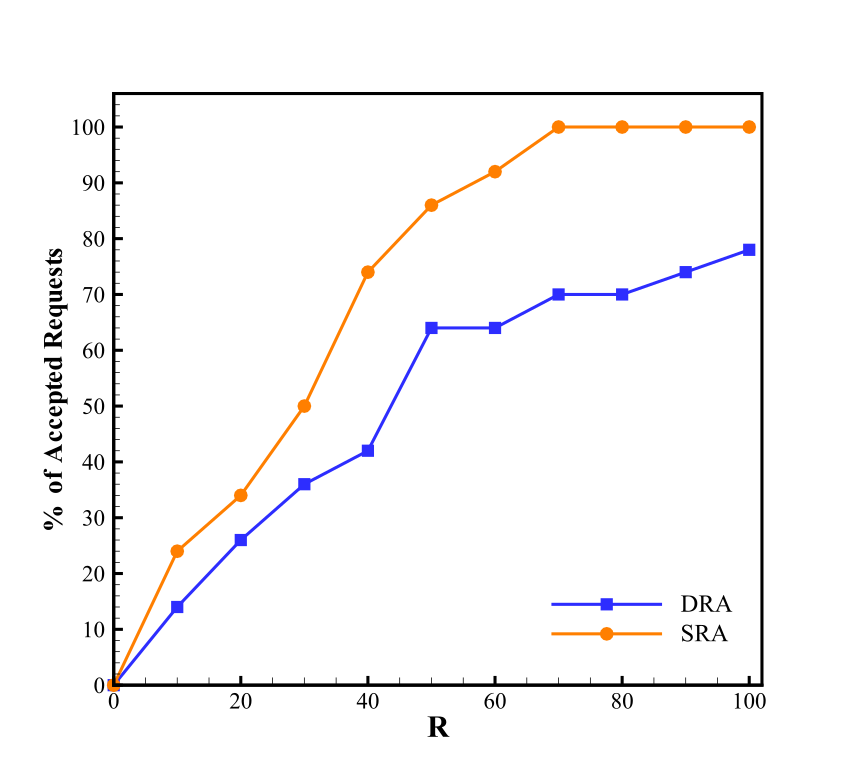}}
\vspace*{-9mm}\hspace*{0mm}\caption{Percentage of Accepted Requests for Constant Requests (b)}
\label{fig2}
\end{figure}

In the following, the number of requests is considered constant and the number of resources available in each time slot is changed. According to Figure \ref{fig2}, as the number of resources increases, the number of requests accepted increases as well. Due to resource sharing between tenants, the SRA algorithm still has a higher number of accepted requests than the DRA algorithm.

\begin{figure}[htbp]
\centerline{\includegraphics[width=1.0\linewidth]{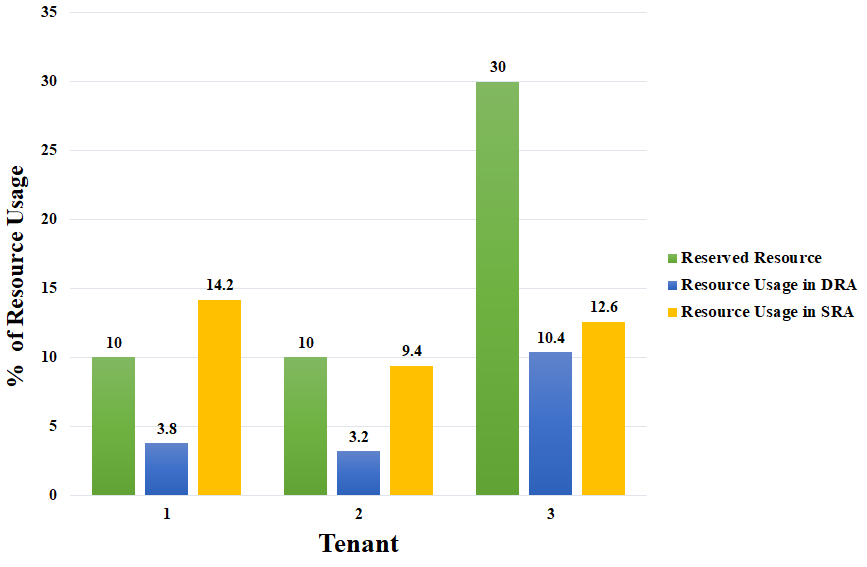}}
\vspace*{-4mm}\hspace*{0mm}\caption{Average of Resource Usage for Tenants ($K=50$, $R=50$)}
\label{fig3}
\end{figure}

In Figure \ref{fig3}, it is shown how the SRA method increases the number of resources used by each tenant compared to the DRA method, given a certain number of requests and resources. In the SRA method, tenants with unused resources share them with others who need them. It is important to remember that resource sharing can violate the reserved resource constraints for each tenant, and this is a negative point. It is necessary to find a compromise between sharing resources and isolating each tenant's reserved resources.

\section{Conclusion}
Slice-Aware calendaring of radio sources in C-RAN is examined in this paper. According to C-RAN's mission, the operator intends to serve a range of slices with varying requirements. In addition, each tenant's resource constraints, delays, and resources should be taken into account. The problem is modeled as an ILP problem, which is NP-Hard with a complex optimal solution. Therefore, two heuristic solutions have been proposed for this problem. In the resource sharing algorithm, the number of accepted requests increases. It should be noted that a compromise must be found between sharing resources and the number of resources reserved for each tenant. The online version of this problem can be reviewed and solved in future work so that all details about the problem aren't available in each time slot.

\bibliographystyle{elsarticle-num}
\bibliography{conference_101719}

\end{document}